\documentclass[aps,prl,twocolumn,longbibliography]{revtex4-1} 

\usepackage{graphicx}
\usepackage[hidelinks]{hyperref}
\usepackage{braket}

\usepackage{bm} 
\usepackage[percent]{overpic} 
\usepackage{ulem} 
\usepackage{xcolor} 
\usepackage{mdframed} 
\newmdenv[backgroundcolor=green!10!white, linecolor=white, leftmargin=10pt, innerleftmargin=5pt,innertopmargin=5pt,innerrightmargin=5pt,innerbottommargin=5pt]{answer}

\newcommand{\unit}[1]{\,\mathrm{#1}}
\renewcommand{\vec}[1]{\boldsymbol{#1}}
\newcommand{\uvec}[1]{\boldsymbol{\hat{#1}}}

\newcommand{\textold}[1]{\textcolor{red!90!black}{\sout{#1}}}
\newcommand{\textnew}[1]{\textcolor{green!50!black}{#1}}

\renewcommand{\textold}[1]{}
\renewcommand{\textnew}[1]{#1}

\usepackage{listings}
\begin{document}
    
    \title{Anisotropic Superfluid Behavior of a Dipolar Bose-Einstein Condensate}
    
    
    \author{Matthias Wenzel}
    \author{Fabian B\"ottcher}
    \author{Jan-Niklas Schmidt}
    \author{Michael Eisenmann}
    \author{Tim Langen}
    \author{Tilman Pfau}
    \author{Igor Ferrier-Barbut}
    \email{i.ferrier-barbut@physik.uni-stuttgart.de}
    
    \affiliation{5{.} Physikalisches Institut and Center for Integrated Quantum Science and Technology (IQST), Universit{\"a}t Stuttgart, Pfaffenwaldring 57, 70569 Stuttgart, Germany}
    
    \date{\today}
    
    \begin{abstract}
        We present transport measurements on a dipolar superfluid using a Bose-Einstein condensate of $^{162}$Dy with strong magnetic dipole-dipole interactions. By moving an attractive laser beam through the condensate we \textnew{observe an anisotropy in superfluid flow. This observation is compatible with an} anisotropic critical velocity for the breakdown of dissipationless flow, which, in the spirit of the Landau criterion, can directly be connected to the anisotropy of the underlying dipolar excitation spectrum. In addition, the heating rate above this critical velocity reflects the same anisotropy.
        Our observations are in excellent agreement with simulations based on the Gross-Pitaevskii equation and highlight the effect of dipolar interactions on macroscopic transport properties, rendering dissipation anisotropic.
    \end{abstract}
    
    
    
    \maketitle
        
    The discovery of superfluidity in liquid helium \cite{Kapitsa1938} is a hallmark of quantum physics at the macroscopic scale.
    The famous Landau criterion \cite{Landau1941} relates the transport properties of a superfluid, namely the maximal velocity for frictionless flow $v_c$, to its spectrum of elementary collective excitations.
    As a consequence, features of the system's excitation spectrum are reflected in the transport properties of the superfluid.
    In the context of ultra-cold atoms, superfluidity and the breakdown thereof have been studied by moving microscopic impurities, i.e. single atoms, which are realized by either stimulated Raman transitions \cite{Chikkatur2000} or with atomic mixtures \cite{Delehaye2015}, allowing a direct comparison to Landau's criterion.
    Other experiments with macroscopic impurities, e.g. laser beams or optical lattices, explored superfluidity in a trapped Bose-Einstein condensate \cite{Raman1999,Onofrio2000a}, a two-dimensional Bose gas \cite{Desbuquois2012} or a Fermi gas in the BEC-BCS crossover regime \cite{Miller2007, Weimer2015}. In \textold{this}\textnew{the latter} case, a reduced critical velocity with respect to the prediction of Landau's criterion is observed.
    
    In the spirit of these pioneering experiments, we perform the first transport measurements on a dipolar Bose-Einstein condensate (dBEC), a superfluid with anisotropic interactions. We observe that the anisotropy of the dispersion relation is reflected in both the anisotropy of the critical velocity and the heating rate above this threshold.
    Our measurements are in excellent agreement with dynamical simulations of the extended Gross-Pitaevskii equation (eGPE) \cite{Wenzel2017} taking into account finite-size effects of the trapped dBEC as well as the characteristics of the moving impurity.
    
    \begin{figure}
        \begin{overpic}[width=0.45\textwidth]{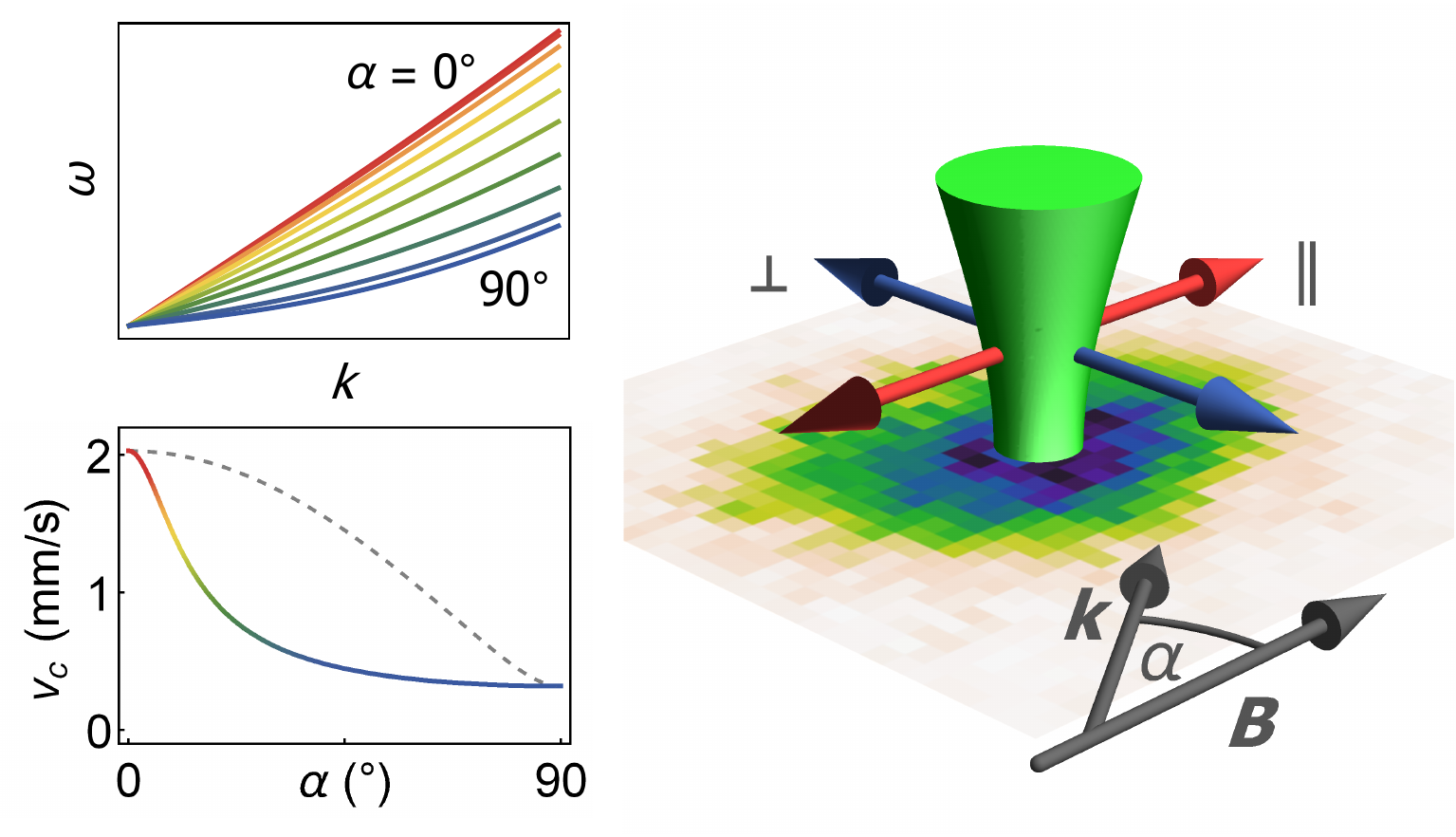}\centering
            \put(-1,52){(a)} \put(-1,26){(b)} \put(45,52){(c)}\end{overpic}
        \vspace{-3mm}
        \caption{\label{fig:1} Probing anisotropic critical velocity.
        (a) Excitation spectrum of a homogeneous dipolar Bose gas. The speed of sound $v_s$ depends on the direction of the excitation $\vec{k}$ with respect to the dipole polarization $\vec{B}$, denoted by the angle $\alpha$. 
        (b) The critical velocity $v_c$ \textold{is given by the speed of sound $v_s = \omega/k$ for $k \rightarrow 0$ and thus becomes anisotropic}\textnew{(solid), as given by eq.~(\ref{eq:vcrit}), becomes anisotropic and is in general lower than $v_s$ (dashed)}.
        (c) Schematic of the experiment. We drag an attractive laser beam through a dipolar condensate perpendicular ($\alpha = 90\unit{^\circ}$, blue) and parallel ($\alpha = 0\unit{^\circ}$, red) to the magnetic field direction.}
    \end{figure}
    
    In order to illustrate the behavior of a dBEC we first focus on the homogeneous gas, where the excitation spectrum
        \begin{equation}\label{eq:disprel}
            \omega(\vec{k}) = k \sqrt{ \textnew{\Big(\frac{\hbar k}{2 m}\Big)^{\!2}} + \frac{g n_0}{m}\Big( 1 + \varepsilon_{dd}\left(3\cos^2\alpha - 1\right) \Big)}
        \end{equation}
    is known analytically \cite{Lahaye2009}. It exhibits an anisotropic dependency on the angle $\alpha$ between excitations with wavevector $\vec{k}$ and the polarization direction $\vec{B}$, see Fig.~\ref{fig:1}a. In a dBEC with a density $n_0$ atoms with mass $m$ are subject to the contact interaction, characterized by the scattering length $a_s$ via $g = 4\pi \hbar^2 a_s/m$, as well as the dipolar interaction, defined by the dipolar length $a_{dd} = \mu_0 \mu_m^2 m / 12\pi\hbar^2$ via the magnetic moment $\mu_m$. The ratio $\varepsilon_{dd} = a_{dd}/ a_s$ of these two length scales describes the relative dipolar strength. The anisotropy of the dipolar excitation spectrum has been confirmed experimentally by Bragg spectroscopy of a chromium dBEC \cite{Bismut2012}.
    
    
    
    The Landau criterion \cite{Landau1941} then relates the anisotropy of the excitation spectrum to the breakdown of superfluidity, since quasiparticles cannot be emitted by an impurity moving at a velocity $v$ smaller than the critical velocity $v_\mathrm{c} = \mathrm{min}\!\left[\,\omega(k)/k \,\right]$ \textnew{in an isotropic fluid. For anisotropic interactions the excitation wavevector $\vec{k}$ does not necessarily coincide with the direction of movement $\vec{\hat{v}}$ of the impurity \cite{Yu2017}. In its generalized form the Landau criterion therefore becomes $v_\mathrm{c} = \mathrm{min}\!\left[\,\omega(\vec{k})/ (\vec{k}\cdot\vec{\hat{v}}) \,\right]$.}
    Applied to the dipolar dispersion relation in eq.~(\ref{eq:disprel}), it yields an anisotropic critical
    velocity
    \begin{equation}\label{eq:vcrit}
        v_c(\alpha) = \bigg( \frac{\sin(\alpha)^2}{v_{s,\perp}^2} + \frac{\cos(\alpha)^2}{v_{s,\parallel}^2} \bigg)^{\!-1/2} \,,
    \end{equation}
    \textold{$old equation$
    which is given by the sound velocity $v_s = \omega(k)/k |_{k \to 0}$ of the homogeneous gas.}
    \textnew{see Fig.~1b (solid line). In general, the acquired $v_c$ is lower than the speed of sound $v_s(\alpha) = \omega(k)/k |_{k \to 0}$ (dashed line) and only coincides with it for a movement parallel $v_{s,\parallel} = v_s(0^\circ)$ or perpendicular $v_{s,\perp} = v_s(90^\circ)$ to the polarization axis.}
    For $^{162}$Dy with a scattering length $a_s = 141(17)\,a_0$ \cite{Tang2018} and a dipolar length $a_{dd} = 131\,a_0$ the critical velocity ranges from $v_{s,\perp} = 0.32\unit{mm/s}$ to $v_{s,\parallel} = 2.0\unit{mm/s}$  for a typical density $n_0 = 10^{20}\unit{m^{-3}}$ as shown in Fig.~1b.
    
    In a confined dipolar system the situation changes since such a system is additionally subject to roton softening \cite{Santos2003,Bisset2013,Chomaz2018} at finite momentum $k$. This collective excitation softening influences $v_c$ \cite{Wilson2010}. In this context the anisotropy of the critical velocity was first predicted in \cite{Ticknor2011} for a quasi-2D dBEC.
    In order to fully account for such confinement-induced effects and other experimental features, full numerical simulations are required.
    
    Here, we perform experiments aimed at measuring the dependence of superfluid flow on the transport direction. Our experimental procedure is as follows, see Fig.~1c. Starting with the setup described in \cite{Schmitt2016} we focus an attractive laser beam ($\lambda = 532\unit{nm}$, along $\uvec{z}$) on a trapped dBEC of $^{162}$Dy atoms. The beam has a waist of $w_0 \approx 1.5\unit{\mu m}$ and power of $P_0 \approx 1.3\unit{\mu W}$. Using the theoretical value of the dynamical polarizability, we estimate the potential depth to $V_0 \approx 0.5\, \mu$, with $\mu$ being the chemical potential of the gas.
    This ``stirring beam'' can be moved transversally over a few $\mathrm{\mu m}$ in the imaging plane by means of two electro-optical deflectors.
    In order to measure superfluid properties we move the beam at a constant velocity $v = 4 \, r_s f_s$  given by the stirring amplitude $r_s$, which is the displacement with respect to the cloud center, and the frequency $f_s$ of a single cycle.
    \textnew{The position $r(t)$ of the stirring beam is thus a triangular periodic function centered around zero with amplitude $r_s$. There is a finite acceleration at the turning points of the triangular motion, leading to the emission of sound waves and thus small heating for velocities below the critical one \cite{Jackson2000}.}
    A minor misalignment leads to a difference in stirring amplitudes for $x$ and $y$, but is fully accounted for as detailed in \cite{SupMat}.
    \textnew{We probe the high-density region avoiding thermal wings by choosing an amplitude $r_s/R_\mathrm{TF} = 0.15 - 0.35$, with $R_\mathrm{TF}$ being the Thomas-Fermi radius of the dBEC.}
    Most of the measurements are carried out in a harmonic trapping potential with frequencies $f_x = 52(1)\unit{Hz} \approx f_y = 49(1)\unit{Hz}$, $f_z = 168(1)\unit{Hz}$ with almost cylindrical symmetry along $\uvec{z}$. In this trap we prepare a dBEC at a condensed fraction of $0.7$ with $1\cdot 10^4$ to $2\cdot 10^4$ atoms in total. Then -- while moving the beam continuously -- the power of the stirring beam is ramped up within $25\unit{ms}$, kept constant for a time $t_\mathrm{stir} = 1\unit{s}$, and ramped down within $25\unit{ms}$ followed by an additional $200\unit{ms}$ for thermalization of the sample. 
    Finally, we extract the temperature $T$ of the sample from in situ images, see \cite{SupMat}.
    \textnew{Due to finite-size effects and experimental noise, our data cannot be considered as a clear proof of superfluidity, but it is in excellent agreement with superfluid flow. To quantify the anisotropy of superfluid flow and compare with simulations}
    \textold{From this data} we extract the critical velocity $v_c$ with a fit function $T(v) = T_0 + h\, t_\mathrm{stir}\, (v/v_c-1) \Theta(v-v_c)$. It is constant in the dissipationless regime below $v_c$ and increases linearly with a given heating rate $\dot T = h\, (v/v_c-1)$ above $v_c$, determined by the heating coefficient $h$.

    In order to take the inhomogenity and finite-size effects of the BEC as well as the finite extent and depth of the stirring beam into account, we conduct dynamic simulations of the extended Gross-Pitaevskii equation \cite{Schmitt2016, Wenzel2017}, which are explained in detail in \cite{SupMat}. The gain in total energy per atom $\Delta E/N$ of a single cycle of the beam's movement is scaled by the number of oscillations $t_\mathrm{stir} f_s$ in the experiment, thus assuming an identical increase in energy induced by the subsequent stirring cycles. Furthermore, there is a non-linear relation between energy and temperature even for the non-interacting Bose gas \cite{Pitaevskii2016}. Since the observed change in temperature is less than 20\%, we assume a linear relation in this regime. Altogether, the simulation data is thus mapped to a temperature $T = T_0 + c \,t_\mathrm{stir} f_s\, \Delta E/ N k_B$ with the Boltzmann constant $k_B$.
    To get agreement between experimental data and simulations we use the coefficient $c$ as a fit parameter scaling only the temperature axis.  \textold{Throughout the data shown in this manuscript}\textnew{For the data presented in Fig.~2, the factor} $c$ is in the range of $0.02 - 0.05$ pointing towards a much weaker heating induced by the subsequent stirring cycles. This parameter also takes into account the mentioned relation between energy and temperature,
    the uncertainty in the potential depth and finite-temperature effects lowering the superfluid fraction \cite{Ghabour2014}. \textold{An accessible finite-temperature theory is lacking for dBECs, but is expected to model the introduced coefficient properly and, even further, would give insights in other unexplored finite-temperature properties.}\textnew{A finite-temperature theory would probably allow to include such effects and further model the introduced coefficient properly.} We emphasize that the rescaling procedure we use here does not influence the critical velocity, which we extract by applying the same fit function as used for the experimental data.
    With the evaluation procedure at hand we now turn to the measurements.
    
    \begin{figure*}
    	\begin{overpic}[width=\textwidth]{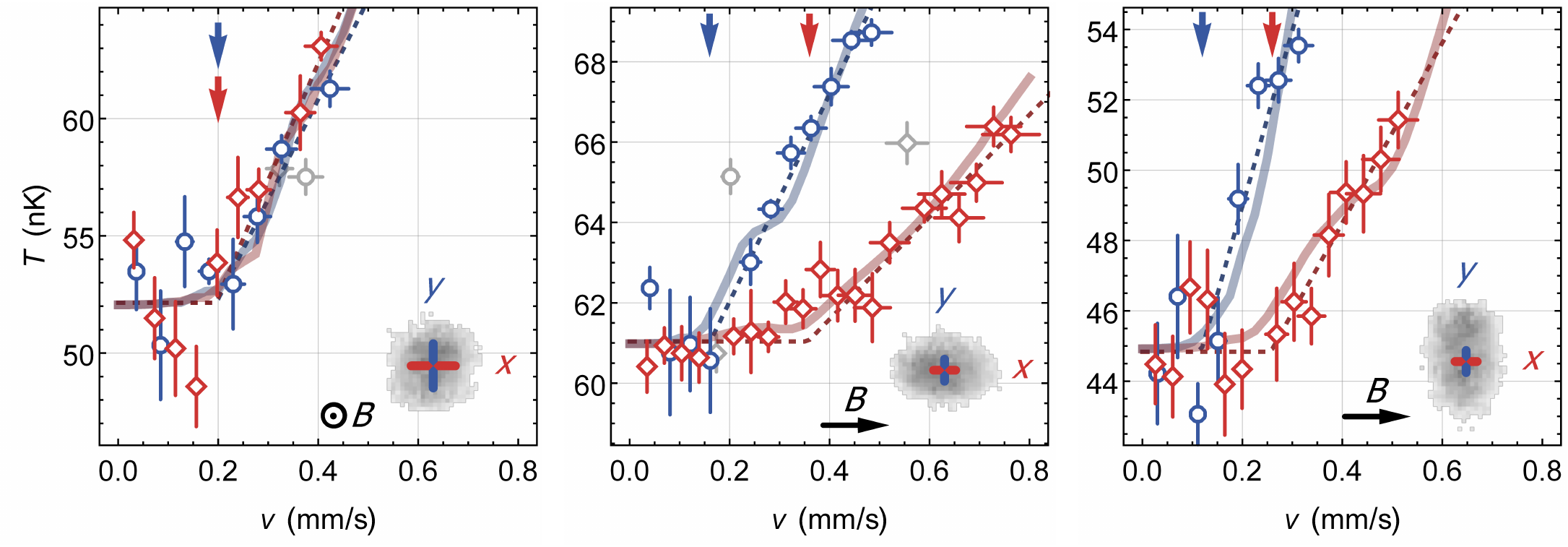}
    		\put(7.4,32){(a)} \put(39.8,32){(b)} \put(72.9,32){(c)} \end{overpic}
        \vspace{-8mm}
        \caption{\label{fig:2} Temperature of the dBEC after stirring for (a) the isotropic case with $\vec{B}\parallel \uvec{z}$ and (b) the anisotropic case with $\vec{B}\parallel\uvec{x}$ \textnew{in an almost cylindrical trap. In (c) the trap is additionally reshaped to invert the cloud aspect ratio.} The stirring beam is moved along  the $x$- (red squares) or $y$- (blue circles) axis, as illustrated in the insets with example in situ images. Critical velocities are extracted by a linear fit (dashed) and marked with arrows. In (a) the response is isotropic with $v_x = 0.20(5)\unit{mm/s}$ and $v_y = 0.20(7)\unit{mm/s}$, while we observe a clear difference in (b) with $v_\perp = 0.16(2)\unit{mm/s}$ along $\uvec{y}$ and $v_\parallel = 0.36(3)\unit{mm/s}$ along $\uvec{x}$. \textnew{In (c) we extract $v_\perp = 0.12(3)\unit{mm/s}$ and $v_\parallel = 0.26(4)\unit{mm/s}$ proving that the observed anisotropy \textnew{remains even} when inverting the anisotropy of the atomic cloud.} Data points with stirring frequency matching the trapping frequencies (gray) are excluded from the analysis. Simulations of the eGPE for a single stirring cycle (solid lines) show excellent agreement with the experiment. See text for further parameters.}
    \end{figure*}
    
    In a first reference measurement we apply the magnetic field $\vec{B} \parallel \uvec{z}$. The problem is therefore isotropic in the $xy$ plane and moving the laser defect along $\uvec{x}$ or $\uvec{y}$ is expected to give the same critical velocity.
    For both stirring directions along $\uvec{x}$ (red diamonds) and $\uvec{y}$ (blue circles) we observe the typical threshold in heating of the dBEC when the velocity of the stirring beam is increased, see Fig.~\ref{fig:2}a. The response is clearly isotropic. Both the critical velocity and the heating coefficient coincide. From the fits (dashed lines) we extract the critical velocity $v_x = 0.20(5)\unit{mm/s}$ and $v_y = 0.20(7)\unit{mm/s}$ with heating coefficients $h_x = 8(5)\unit{nK/s}$ and $h_y = 9(8)\unit{nK/s}$, respectively.
    For this measurement the stirring frequency is varied between $f_s = 3$ and $60\unit{Hz}$. Data points with stirring frequency at the transversal trap frequency (gray) are excluded from the analysis, since the coupling to the center-of-mass oscillation in the trap can influence the energy transfer.
    The agreement with the simulation data (solid lines) is remarkable. We extract a critical velocity of $v_{x,\,\mathrm{sim}} = v_{y,\,\mathrm{sim}} = 0.21(1)\unit{mm/s}$ in excellent agreement with the presented experimental values.
    
    We now turn to the anisotropic case with the magnetic field $\vec{B} \parallel \uvec{x}$ pointing along one of the stirring directions. Due to magnetostriction the cloud is deformed to an aspect ratio of $\kappa = R_x/R_y = 1.4$ \cite{Lahaye2009} in the imaging plane with Thomas-Fermi radii $R_x = 6.0\unit{\mu m}$ and $R_y = 4.3\unit{\mu m}$. In this configuration the cloud is elongated along the magnetic field, thus the mean dipolar interaction is predominantly attractive and therefore the peak density $n_0 = 1.7\cdot10^{20}\unit{m^{-3}}$ is a factor of two higher as compared to the previous case.
    More importantly, the dispersion relation becomes anisotropic when comparing the stirring directions along $\uvec{x} \parallel \vec{B}$ and $\uvec{y} \perp \vec{B}$. In consequence, we directly observe a factor of two difference in critical velocity, as shown in Fig.~2b. The extracted values are $v_\perp = 0.16(2)\unit{mm/s}$ and $v_\parallel = 0.36(3)\unit{mm/s}$ with heating coefficients $h_\perp = 4.2(9)\unit{nK/s}$ and $h_\parallel= 4.5(9)\unit{nK/s}$, that agree within the experimental error. The difference in heating rates $\dot{T} = h\, (v/v_c-1)$, as given by the slope in the figure, can thus be fully attributed to the anisotropy of the critical velocity. From this fact we infer that the anisotropy in both critical velocity and heating rate share a common cause in the anisotropy of collective excitations. Comparing to simulation data yields excellent agreement, as can be seen in Fig.~2b. We stress that a single fit parameter $c$ is used for both curves. The anisotropy in heating rate is thus very well reproduced by the simulation. We further extract $v_{\perp,\,\mathrm{sim}} = 0.16(1)\unit{mm/s}$ and $v_{\parallel,\,\mathrm{sim}} = 0.35(2)\unit{mm/s}$ in excellent agreement with the experiment.
    
    
    In order to ensure that the observed anisotropy is not trivially caused by the anisotropic cloud shape, we invert the aspect ratio of the cloud to $\kappa = R_x/R_y \approx 1.4^{-1}$ by adjusting the trapping potential. The trap frequencies in this case are $\{f_x, f_y, f_z\} = \{81(2), 39(1), 140(1)\}\unit{Hz}$ counteracting the magnetostriction along the magnetic field axis $\vec{B} \parallel \uvec{x}$. This leads to measured sizes $R_x = 4.3\unit{\mu m}$ and $R_y = 5.8\unit{\mu m}$ of the condensate. \textold{Figure~3 shows the corresponding data along with the previous measurement for comparison.} The extracted critical velocities are $v_\perp = 0.12(3)\unit{mm/s}$ and $v_\parallel = 0.26(4)\unit{mm/s}$ again with compatible heating coefficients $h_\perp = 6(3)\unit{nK/s}$ and $h_\parallel = 7(3)\unit{nK/s}$\textnew{, as shown in Fig.~2c}.
    The observed anisotropy of transport remains in the same direction even though the cloud aspect ratio was inverted, providing conclusive evidence that it arises directly from the dipolar anisotropy. Once again for this data set, excellent agreement with simulations is found.
    
    We further compare the measured $v_c$ to the speed of sound $v_s$ of the homogeneous dipolar gas introduced in eq.~(\ref{eq:vcrit}). For the given peak density the latter is $v_{s,\perp} = 0.42\unit{mm/s}$ and $v_{s,\parallel} = 2.6\unit{mm/s}$, respectively. The measured critical velocity $v_c/v_s = 0.1 - 0.4$ is thus well below the expected speed of sound in the cloud center. This value is in agreement with the critical velocity measured in the pioneering experiment with a contact-interacting BEC \cite{Onofrio2000a}. An obvious effect lowering the measured $v_c$ is the inhomogeneuos density distribution both along the beam and transversally \cite{Fedichev2001}. Vortex formation is a dominant effect for repulsive obstacles lowering the density, but should be supressed in our experiment with an attractive beam \cite{Singh2016}. Yet, the macrosocopic size of the beam can influence the measured critical velocity as well \cite{Stiessberger2000}. 
    
    
    In conclusion, we performed the first transport measurements on a dipolar BEC. The strong dipole-dipole interaction of dysprosium atoms renders the excitation spectrum of the dBEC, and thus the critical velocity for the breakdown of superfluidity, anisotropic. We investigate the latter by measuring the heating caused by moving an attractive laser beam through the condensate. We find excellent agreement comparing our data taken at a sizeable thermal fraction to dynamic simulations of the eGPE, which is a zero temperature theory. We therefore deduce that the effect of thermal excitations has a negligible influence on the critical velocity in our experiment.
    \textnew{As discussed earlier, roton softening of the excitation spectrum can decrease the critical velocity \cite{Wilson2010}. Yet, for the current set of experiments with dipolar strength of $\varepsilon_{dd} < 1$ in conjunction with a weak confinement along the magnetic field this effect is likely negligible. Increasing both quantities could lead to an observable reduction of the critical velocity, which is an interesting perspective for future studies.}
    An anisotropic dispersion relation is expected to have many more implications on hallmark properties of superfluids, e.g.~on vortices in rotating systems. In future experiments we expect to find an anisotropic density distribution around a vortex core \cite{Yi2006}. Furthermore, this effect induces anisotropic vortex-vortex interactions \cite{Mulkerin2013} leading to transitions between vortex lattices of different symmetries \cite{Cooper2005}.
    
    \begin{acknowledgments}
        We thank A{.} Pelster and A. Bala\v{z} for valuable discussions as well as Z.-Q. Yu for pointing us to the generalized form of the Landau criterion. This work is supported by the German Research Foundation (DFG) within FOR2247 under Pf381/16-1, Pf381/20-1, and HBFG INST41/1056-1. IFB and TL acknowledge support from the EU within Horizon2020 Marie Sk{\l}odowska Curie IF (703419 DipInQuantum and 746525 coolDips, respectively). TL acknowledges support from the Alexander von Humboldt Foundation through a Feodor Lynen Fellowship.
    \end{acknowledgments}

    
    \bibliography{paper}

    
    \vspace{3cm}
    \pagebreak
    \onecolumngrid
    \begin{center}
    	{\Large \textbf{Supplemental Material}}
    \end{center}
    \twocolumngrid
    \vspace{1cm}

    \setcounter{equation}{0}
    \setcounter{figure}{0}
    \setcounter{table}{0}
    \setcounter{page}{1}
    \makeatletter
    \renewcommand{\theequation}{S\arabic{equation}}
    \renewcommand{\thefigure}{S\arabic{figure}}

        \subsection{Imaging \& Temperature Extraction}

        After the stirring sequence (described in the main text) and subsequent thermalization we ramp up the magnetic field $B_z \approx 10\unit{G}$ in $100\unit{\mu s}$ since the phase-contrast imaging scheme relies on a magnetic field parallel to the imaging beam along $z$. The latter is detuned by $20\,\Gamma$ with respect to the $421\unit{nm}$ transition, see \cite{Kadau2016}. From the acquired in situ images we extract the condensed fraction $N_0/N$ by fitting a thermal Gaussian distribution plus a Thomas-Fermi parabola. The temperature is then extracted from the relation $N_0/N = 1 - (T/T_c)^{3}$ with a critical temperature $T_c$ calculated including finite-size effects \cite{Pitaevskii2016} and interactions \cite{Glaum2007}. For the presented measurements $T_c$ ranges between $59$ and $77\unit{nK}$.
        
        \subsection{Velocity Calibration}
        
        The displayed velocity $v_s = 4 r_s f_s$ of the stirring beam depends linearly on the calibration of the stirring displacement $r_s$ for the two directions.
        To reduce systematics we calibrate the magnification of the imaging system in a first step. For this purpose we move the objective, which is mounted on a piezo-stage. From the resulting relative displacement of the cloud on the camera we extract a magnification of $M_x = M_y = 44.2(1)$, measured for both $x$ and $y$ direction independently. We further confirm with a raytracing software that the lateral displacement of the objective over $< 100\unit{\mu m}$ should not affect the magnification of the imaging system due to, e.g. imaging aberrations.
        
        In a second step we load all atoms in the stirring beam with larger power and move it over the full range $d = 2\, r_\mathrm{max}$ of the electro-optical deflector system along one direction and take images at various positions. By a linear fit to the position data we extract the stirring amplitudes $r_{\mathrm{max},x} = 2.9(2)\unit{\mu m}$ and $r_{\mathrm{max},y} = 3.4(2)\unit{\mu m}$.
        We attribute the difference in $r_\mathrm{max}$ to the electro-optical deflector system in conjunction with a slight misalignment of the focus position of the stirring beam. To deflect the beam along $x$ and $y$ two $20\unit{cm}$ deflector tubes are placed consecutively along the collimated beam. This offset in deflector position causes an amplitude aspect ratio $r_{\mathrm{max},y} / r_{\mathrm{max},x}$ that varies between $0.89$ and $1.17$ within the Rayleigh range $z_R = \pm7.5\unit{\mu m}$ of the stirrer's focus. A minor misalignment therefore leads to the measured difference in stirring amplitude.
        This way we characterize the stirring amplitudes in both $x$ and $y$ direction and rescale the velocity appropriately, taking the mentioned misalignment into account. In order to probe only the high-density region of the condensate, we restrict the measurement to $r_s/r_\mathrm{max} = 0.6$ for the first and $0.3$ for the second and third one. The smaller radius is used, because the size orthogonal to the magnetic field is smaller due to magnetostriction. For these measurements we therefore probe a higher frequency range $f_s = 10 - 220\unit{Hz}$.
        
        Yet the extracted Thomas-Fermi radius is prone to imaging aberrations. This effect can be seen for the first data set with $B\parallel z$ yielding an aspect ratio of $R_x/R_y = 5.1 / 5.8 \approx 0.88$.
        Using a variational ansatz \cite{Lahaye2009} we estimate the expected aspect ratio in the described trap to $R_x / R_y = 5.4/5.8 \approx 0.93$ and attribute the residual $6\%$ difference to aberrations of the imaging system. Since these can influence the calibration of the velocity, as outlined above, we quadratically add this error to the one in $r_\mathrm{max}$, which then yields the error of the displayed velocity $v_s$.

        \subsection{Simulations}
        
        The presented numerical simulations are based on the extended Gross-Pitaevskii equation (eGPE) as outlined in \cite{Schmitt2016, Wenzel2017}. The correction due to beyond mean-field effects is negligible for the simulations presented here.
        We solve the eGPE on a 3D grid using the split-step method with a Crank-Nicolson scheme for the derivatives.
        For the external potential we implement a time-dependent attractive ``stirrer'' $V_\mathrm{stir}(\vec{r},t)$ in addition to the static harmonic trap $V_\mathrm{trap}(\vec{r})$.
        The former corresponds to a gaussian beam along $z$ with waist size $w_0 = 1.5\unit{\mu m}$, power of $P = 1\unit{\mu W}$ and calculated polarizability $\mathrm{Re\{\alpha\} = 429\unit{ a. u.}}$ matching the conditions of the experiment.
        First, we solve in imaginary-time evolution in order to prepare the condensate ground state in this combined potential. In a second step we propagate the wave function in real-time evolution, moving the stirring potential with a constant velocity $v_s$ in the desired direction from $r = 0$ to $r_s$, then to $- r_s$, and back to $r = 0$.
        Finally, we determine the induced increase in energy per atom $\Delta E/N$ by comparing the system's total energy before and after such a stirring cycle. The temperature is then extracted as described in the main text.
        The scaling coefficient $c$ is $0.022$, $0.0375$ and $0.05$ for the first, second and third data set, respectively, resulting from fits to the experimental data.
        \\
        We note that preparing the ground state in the harmonic trap only and subsequent adiabatic ramping of the stirring potential's depth leads to similar results, having the disadvantage of longer simulation times.
        \\


    %

\end{document}